\documentclass[twocolumn,floatfix,times]{aastex63}

\usepackage{graphicx}
\usepackage{longtable}\LTcapwidth=\textwidth
\usepackage{multirow}
\usepackage{subfigure}
\usepackage{ulem}
\usepackage{comment}

\newcommand\chem[1]{\ensuremath{\mathrm{#1}}}
\newcommand{\cmq}{cm$^{-3}$}

\newcommand{\kms}{\mbox{km s$^{-1}$}}
\newcommand{\ntdp}{N$_2$D$^+$}
\newcommand{\dv}{\mbox {$\Delta v$}}
\newcommand{\tk}{\mbox{$T_K$}}
\newcommand\ee[1]{\ensuremath{\times 10^{#1}}}
\newcommand{\msun}{M_\odot}

\received{XXXX, 2020}
\revised{XXXX, 2020}
\accepted{\today}
\submitjournal{ApJ}

\begin{document}

\title{ALMA Survey of Orion Planck Galactic Cold Clumps (ALMASOP): Detection of extremely high density  compact structure of prestellar cores and multiple substructures within }

 \correspondingauthor{Dipen Sahu; Tie Liu; Sheng-Yuan Liu}
 \email{dsahu@asiaa.sinica.edu.tw; liutie@shao.ac.cn}

\author[0000-0002-4393-3463]{Dipen Sahu}
\affiliation{Academia Sinica Institute of Astronomy and Astrophysics, 11F of AS/NTU Astronomy-Mathematics Building, No.1, Sec. 4, Roosevelt Rd, Taipei 10617, Taiwan, R.O.C.}

\author[0000-0012-3245-1234]{Sheng-Yuan Liu}
\affiliation{Academia Sinica Institute of Astronomy and Astrophysics, 11F of AS/NTU Astronomy-Mathematics Building, No.1, Sec. 4, Roosevelt Rd, Taipei 10617, Taiwan, R.O.C.}

\author[0000-0002-5286-2564]{Tie Liu}
\affiliation{Shanghai Astronomical Observatory, Chinese Academy of Sciences, 80 Nandan Road, Shanghai 200030, People's Republic of China}

\author[0000-0001-5175-1777]{Neal J. Evans II}
\affiliation{Department of Astronomy The University of Texas at Austin 2515 Speedway, Stop C1400 Austin, TX 78712-1205, USA}

\author[0000-0001-9304-7884]{Naomi Hirano}
\affiliation{Academia Sinica Institute of Astronomy and Astrophysics, 11F of AS/NTU Astronomy-Mathematics Building, No.1, Sec. 4, Roosevelt Rd, Taipei 10617, Taiwan, R.O.C.}

\author[0000-0002-8149-8546]{Ken'ichi Tatematsu}
\affiliation{Nobeyama Radio Observatory, National Astronomical Observatory of Japan,
National Institutes of Natural Sciences,
462-2 Nobeyama, Minamimaki, Minamisaku, Nagano 384-1305, Japan}

\author[0000-0002-3024-5864]{Chin-Fei Lee}
\affiliation{Academia Sinica Institute of Astronomy and Astrophysics, 11F of AS/NTU Astronomy-Mathematics Building, No.1, Sec. 4, Roosevelt Rd, Taipei 10617, Taiwan, R.O.C.}

\author[0000-0003-2412-7092]{Kee-Tae Kim}
\affiliation{Korea Astronomy and Space Science Institute, 776 Daedeokdae-ro, Yuseong-gu, Daejeon 34055, Republic of Korea }
\affiliation{University of Science and Technology, Korea (UST), 217 Gajeong-ro, Yuseong-gu, Daejeon 34113, Republic of Korea}

\author[0000-0002-2338-4583]{Somnath Dutta}
\affiliation{Academia Sinica Institute of Astronomy and Astrophysics, 11F of AS/NTU Astronomy-Mathematics Building, No.1, Sec. 4, Roosevelt Rd, Taipei 10617, Taiwan, R.O.C.}


  
\author{Dana Alina}
\affiliation{Department  of  Physics,  School  of  Sciences  and  Humanities, Nazarbayev University, Nur-Sultan 010000, Kazakhstan}

\author[0000-0002-9574-8454]{Leonardo Bronfman}
\affiliation{Departamento de Astronomía, Universidad de Chile, Camino el Observatorio 1515, Las Condes, Santiago, Chile}

\author{Maria Cunningham}
\affiliation{School of Physics, University of New South Wales(UNSW), Sydney, NSW 2052, Australia.}

\author[0000-0002-5881-3229]{David J. Eden}
\affiliation{Astrophysics Research Institute, Liverpool John Moores University, iC2, Liverpool Science Park, 146 Brownlow Hill, Liverpool, L3 5RF, UK.}
 
 \author{Guido Garay}
\affiliation{Departamento de Astronomía, Universidad de Chile, Camino el Observatorio 1515, Las Condes, Santiago, Chile}

\author{Paul F. Goldsmith}
\affiliation{Jet Propulsion Laboratory, California Institute of Technology, 4800 Oak Grove Drive, Pasadena, CA 91109, USA}

\author[0000-0002-3938-4393]{Jinhua He}
\affiliation{Yunnan Observatories, Chinese Academy of Sciences, 396 Yangfangwang, Guandu District, Kunming, 650216, P. R. China}
\affiliation{Chinese Academy of Sciences South America Center for Astronomy, National Astronomical Observatories, CAS, Beijing 100101, China}
\affiliation{Departamento de Astronom\'{i}a, Universidad de Chile, Casilla 36-D, Santiago, Chile}
 
 \author[0000-0002-1369-1563]{Shih-Ying Hsu}
\affiliation{National Taiwan University (NTU), Taiwan, R.O.C.}
\affiliation{Academia Sinica Institute of Astronomy and Astrophysics, 11F of AS/NTU Astronomy-Mathematics Building, No.1, Sec. 4, Roosevelt Rd, Taipei 10617, Taiwan, R.O.C.}
 
 \author[0000-0003-2069-1403]{Kai-Syun Jhan}
\affiliation{National Taiwan University (NTU), Taiwan, R.O.C.}
\affiliation{Academia Sinica Institute of Astronomy and Astrophysics, 11F of AS/NTU Astronomy-Mathematics Building, No.1, Sec. 4, Roosevelt Rd, Taipei 10617, Taiwan, R.O.C.}
 
 \author[0000-0002-6773-459X]{Doug Johnstone}
\affiliation{NRC Herzberg Astronomy and Astrophysics, 5071 West Saanich Rd, Victoria, BC, V9E 2E7, Canada}
\affiliation{Department of Physics and Astronomy, University of Victoria, Victoria, BC, V8P 5C2, Canada}

\author[0000-0002-5809-4834]{Mika Juvela}
\affiliation{Department of Physics, P.O.Box 64, FI-00014, University of Helsinki, Finland}
 
 \author[0000-0003-2011-8172]{Gwanjeong Kim}
\affil{Nobeyama Radio Observatory, National Astronomical Observatory of Japan, 
National Institutes of Natural Sciences, 
462-2 Nobeyama, Minamimaki, Minamisaku, Nagano 384-1305, Japan}
 
 \author{Yi-Jehng Kuan}
\affiliation{Department of Earth Sciences, National Taiwan Normal University, Taipei, Taiwan (R.O.C.)}
\affiliation{Academia Sinica Institute of Astronomy and Astrophysics, 11F of AS/NTU Astronomy-Mathematics Building, No.1, Sec. 4, Roosevelt Rd, Taipei 10617, Taiwan, R.O.C.}

 \author[0000-0003-4022-4132]{Woojin Kwon}
\affiliation{Department of Earth Science Education, Seoul National University, 1 Gwanak-ro, Gwanak-gu, Seoul 08826, Republic of Korea}

\author[0000-0002-3179-6334]{Chang Won Lee}
\affiliation{Korea Astronomy and Space Science Institute, 776 Daedeokdae-ro, Yuseong-gu, Daejeon 34055, Republic of Korea }
\affiliation{University of Science and Technology, Korea (UST), 217 Gajeong-ro, Yuseong-gu, Daejeon 34113, Republic of Korea}

\author[0000-0003-3119-2087]{Jeong-Eun Lee}
\affiliation{School of Space Research, Kyung Hee University, 1732, Deogyeong-Daero, Giheung-gu Yongin-shi, Gyunggi-do 17104, Korea}

 \author[0000-0003-3010-7661]{Di Li}
\affiliation{National Astronomical Observatories, Chinese Academy of Sciences, Beijing 100101, China} 
\affiliation{NAOC-UKZN Computational Astrophysics Centre, University of KwaZulu-Natal, Durban 4000, South Africa}

\author{ Pak Shing Li}
\affiliation{Department of Astronomy, University of California, Berkeley, CA 94720, USA}
 
 \author[0000-0003-1275-5251]{Shanghuo Li}
\affiliation{Korea Astronomy and Space Science Institute, 776 Daedeokdae-ro, Yuseong-gu, Daejeon 34055, Republic of Korea }

\author{Qiu-Yi Luo}
\affiliation{Shanghai Astronomical Observatory, Chinese Academy of Sciences, 80 Nandan Road, Shanghai 200030, People's Republic of China}

\author{Julien Montillaud}
\affiliation{Institut UTINAM – UMR 6213 – CNRS – Univ. Bourgogne
Franche Comt\'e, OSU THETA, 41bis avenue de l'Observatoire,
25000 Besan\c{c}on, France}

 \author{Anthony Moraghan}
 \affiliation{Academia Sinica Institute of Astronomy and Astrophysics, 11F of AS/NTU Astronomy-Mathematics Building, No.1, Sec. 4, Roosevelt Rd, Taipei 10617, Taiwan, R.O.C.}

 \author{Veli-Matti Pelkonen}
\affiliation{Institut de Ci\`{e}ncies del Cosmos, Universitat de Barcelona, IEEC-UB, Mart\'{i} i Franqu\'{e}s 1, E08028 Barcelona, Spain}

\author{Sheng-Li Qin}
\affiliation{Department of Astronomy, Yunnan University, and Key Laboratory of Particle Astrophysics of Yunnan Province, Kunming, 650091, People's Republic of China}

\author{Isabelle Ristorcelli}
 \affiliation{ Universit\'e de Toulouse, UPS-OMP, IRAP, F-31028 Toulouse cedex 4, France}
 
 \author[0000-0002-7125-7685]{Patricio Sanhueza}
\affiliation{National Astronomical Observatory of Japan, National Institutes of Natural Sciences, 2-21-1 Osawa, Mitaka, Tokyo 181-8588, Japan}
\affiliation{Department of Astronomical Science, SOKENDAI (The Graduate University for Advanced Studies),
2-21-1 Osawa, Mitaka, Tokyo 181-8588, Japan}

\author[0000-0001-8385-9838]{Hsien Shang}
\affiliation{Academia Sinica Institute of Astronomy and Astrophysics, 11F of AS/NTU Astronomy-Mathematics Building, No.1, Sec. 4, Roosevelt Rd, Taipei 10617, Taiwan, R.O.C.}

\author{Zhi-Qiang Shen}
\affiliation{Shanghai Astronomical Observatory, Chinese Academy of Sciences, 80 Nandan Road, Shanghai 200030, People's Republic of China}
 
\author[0000-0002-6386-2906]{Archana Soam}
\affil{SOFIA Science Center, Universities Space Research Association, NASA Ames Research Center, Moffett Field, California 94035, USA}

\author{Yuefang Wu}
\affiliation{Department of Astronomy, Peking University, 100871 Beijing, People's Republic of China}

\author[0000-0003-2384-6589]{Qizhou Zhang}
\affiliation{Center for Astrophysics | Harvard \& Smithsonian, 60 Garden Street, Cambridge, MA 02138, USA}

 \author[0000-0003-0356-818X]{Jianjun Zhou}
 \affiliation{ Xinjiang Astronomical Observatory, CAS: Urumqi, Xinjiang, CN}




\begin{abstract}

Prestellar cores are self-gravitating dense and cold structures within molecular clouds where future stars are born.
They are expected, at the stage  of  transitioning to the protostellar phase, to harbor
centrally concentrated dense (sub)structures that will seed the formation of a new star or the
binary/multiple stellar systems.
Characterizing this critical stage of evolution is key to our understanding of star formation. 
In this work, we report the detection of high density (sub)structures  on the  thousand-au scale in a sample of dense prestellar cores. 
Through our recent ALMA observations towards the Orion molecular cloud, we have found five extremely dense prestellar cores, which have centrally concentrated regions $\sim$ 2000 au in size, and several $10^7$ \cmq\ in average density.   Masses of these centrally dense regions are in the range of 0.30 to 6.89 M$_\odot$. 
{\it For the first time}, our higher resolution observations (0.8$'' \sim $ 320 au) further reveal that one of the cores shows clear signatures of fragmentation; such individual substructures/fragments have sizes of 800 -1700 au, masses of 0.08 to 0.84 M$_\odot$, densities of $2 - 8\times 10^7$ \cmq, and separations of $\sim 1200$ au. 
The substructures are massive enough ($\gtrsim 0.1~M_\odot$) to form young stellar objects and are likely examples of the earliest stage of stellar embryos which can lead to widely ($\sim$ 1200 au) separated multiple systems. 

\end{abstract}


\section{Introduction }

Stars form from the gravitational collapse of dense molecular cloud cores in the interstellar medium. 
Characterization of the dense cores is therefore of great importance for gaining insights into the initial conditions and evolutionary stages in star formation. Starless cores, which are condensations of matter without any embedded young stellar objects (YSOs), are considered as the earliest phase of star formation. A subset of the starless cores, called prestellar cores, are gravitationally bound and will presumably collapse to form YSOs \citep{andre2014}. The evolution of prestellar cores to YSOs, and whether they form single or multiple stellar systems, however, is far from being understood.

 While extensive efforts have been made previously to investigate the physical and chemical properties of starless and prestellar cores with single element telescopes \citep[e.g.,][]{Caselli2011}, the densest and presumably innermost 1000~au of such cores are yet to be characterized due to the limited angular, hence spatial, resolution.
This scale is however crucial for testing theories of fragmentation to form multiple systems.

To this end,
survey observations of starless cores in nearby molecular clouds (including, for example, Perseus, Ophiuchus, and Chamaeleon) with modern (sub)millimeter interferometers, such as the Combined Array for Research in Millimeter-wave Astronomy (CARMA), the Submillimeter Array (SMA), and the Atacama Large Millimeter and submillimeter Array (ALMA) were conducted \citep[e.g][]{Schnee2010,Schnee2012a,Dunham2016,Kirk2017}.
In these survey observations,  very few localized sources were recovered 
because of the insensitivity of interferometers to small fluctuations in extended emission.
Among those detected sources, some turned out to be protostellar, while others remained starless  with no additional substructure in their (sub)millimeter continuum emission.
Such results suggest that the density profiles of starless cores are 
predominantly flat,
consistent with the earlier suggestions \citep[e.g.,][]{Ward-Thompson1994,Ward-Thompson1999,Shirley2000,Caselli2019}.
A profile having nearly constant density 
at small radii ($\sim$ few 1000 au)
is reminiscent of the ``flat zone" in Bonnor-Ebert spheres \citep{Ebert1955,Bonnor1956}.
 
Meanwhile, observations indicate that the multiplicity fraction and the companion star fraction are highest in Class 0 protostars and decrease in more evolved protostars \citep{chen2013,Tobin2016}.  This implies that multiple systems may  develop at an even earlier phase.
Indeed, some theoretical works suggest that substructures in prestellar cores can be produced by turbulent fragmentation at a scale of 1000 au, which will form wide-multiple stellar systems \citep{Offner2010}, and these  should be visible using  sensitive interferometers \citep{Offner2012}. 
However, none of the observational surveys described in the  previous paragraph found secondary/multiple substructures within dense cores at a scale of 1000 au. 
 Although a few studies report cases of substructure and fragmentation in specific starless cores  \citep[e.g.,][]{Kirk2009,chen2010,Nakamura2012,Takahashi2013,Pineda2015,Friesen2014,ohashi2018,Tatematsu2020}
they are associated with  
 high-density condensation, 
and not related to the substructures (1000 au scale)  in the central dense region of prestellar cores.  

 In a recent ALMA survey, \citet{Tokuda2020} reported the presence of very low mass ($\sim 10^{-2}$ M$_\odot$ , 
$n_{\mbox{\scriptsize H$_2$}}$ $\sim 10^{5}~$\cmq) substructure toward starless cores at a scale of 1000 au; these substructures are not 
 massive enough  to produce  multiple stars.  In another observational effort to study the inner dense region of the prestellar core L1544 \citep{Caselli2019}, 
no substructures were detected inside the central dense region ($\sim 2000$ au with $n_{\mbox{\scriptsize H$_2$}}$ $\geq 10^{6}$\cmq) of the core, where stellar multiplets are expected to potentially develop.

Recently, we performed a survey  \citep[ALMA Survey of Orion Planck Galactic Cold Clumps or ALMASOP;][]{Dutta2020} with ALMA, in which we targeted 72 cores toward the Orion Molecular Clouds \citep[distance $\sim $ 400 pc;][]{Kounkel2018}.
An overview of the ALMASOP is presented in \citet{Dutta2020}. Among the 72 targets in ALMASOP observations,  23 were previously classified as (candidate) starless cores. 
While all these  starless cores were detected in the SCUBA-2 observations (with a core scale of $\sim 0.1$ pc or 20000 au), only 16 of them are detected by the ALMASOP ACA observations (scale $\sim 0.03$ pc or 6000 au). 
 In this letter, we highlight the detection of a central compact dense  structure toward a sample of five cores and  the discovery of substructures within one of them using the ALMA 12-m array.

\section{Observations\label{sec:Obs}}

The observations of ALMASOP (project ID:2018.1.00302.S. ; PI: Tie Liu) were carried out with ALMA in Cycle 6 toward 72 fields during 2018 October to 2019 January. 
The observations were executed in four blocks in three different array configurations: 12m C43-5 (TM1), 12m C43-2 (TM2), and 7m ACA,  resulting in a resolution ranging 0\farcs34 - 5.5\arcsec in the 1.3 mm band.
 The correlator was configured into four spectral windows with 1.875 GHz  bandwidth, which provides a spectral resolution of 1.129 MHz,
corresponding to velocity resolutions between 1.465 and 1.563 \kms.
We adopted this coarse velocity resolution to facilitate efficient continuum observations and to maximize the spectral line coverage.
The spectral set-up covers the continuum emission at 233.0 GHz and 216.6 GHz, and  offers 
simultaneous coverage of the molecular lines CO (2-1), \chem{C^{18}O} (2-1), \chem{N_2D^+} (3-2), SiO (5-4) and other hot corino tracers as well  \citep[see][]{Hsu2020}.  
The remaining details of the observational parameters are presented in \citet{Dutta2020}.

The calibrated visibility data were obtained using CASA 5.4  \citep[Common Astronomy Software Applications package;][]{McMullin2007} pipeline script as delivered by the observatory.  The visibility data for different configurations and executions corresponding to the 72 sources were then separated into continuum and spectral data,  and  imaged jointly. The 1.3~mm continuum images of the sources are generated through CASA’s tclean task with the `automask’ on, the hogbom deconvolver, and a robust weighting of 0.5.

\begin{figure*}[ht!]
    \centering
    \includegraphics[width=18cm]{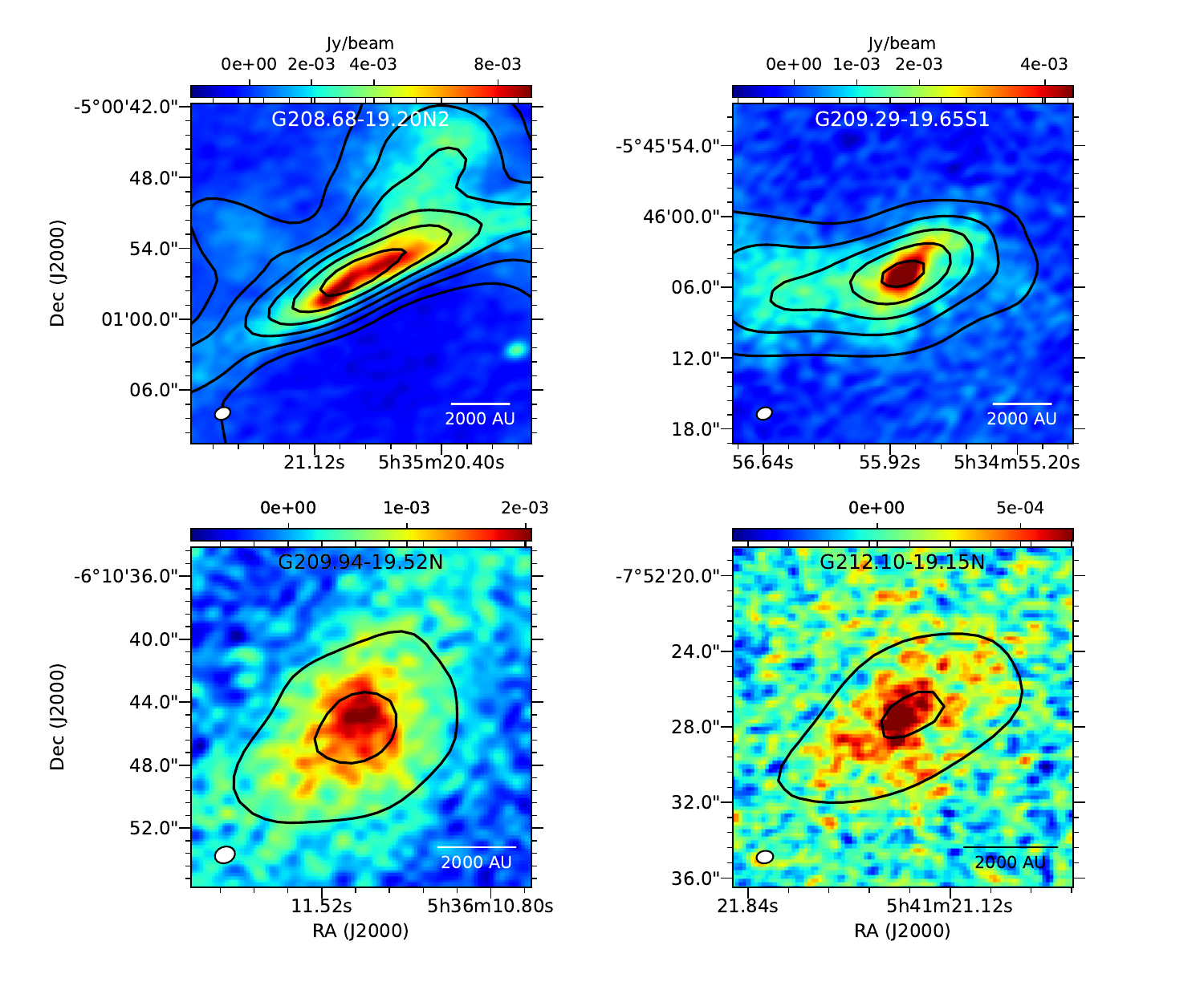}
    \caption{The central dense region of the prestellar cores G208.68-19.20N2 (top-left), G209.29-19.65S1 (top-right), G209.94-19.52N (bottom-left), and G212.10-19.15N1 (bottom-right). The contours represent the 1.3 mm continuum observed by the ACA with levels 5,10,15$\sigma$..., where $\sigma$=3.4, 1.9, 2.5, 1.6~mJy/beam, respectively for the sources; and the color scales show the ALMA TM2 detection of the centrally dense region. 
 Beam sizes  (typically 1.2")  for each source are shown in the lower left of each panel.}
    \label{fig1}
\end{figure*}

\section{Results}


\subsection{Detection of  compact dense structure  inside prestellar cores}

\begin{figure*}[ht!]
    \centering
    \includegraphics[width=18cm]{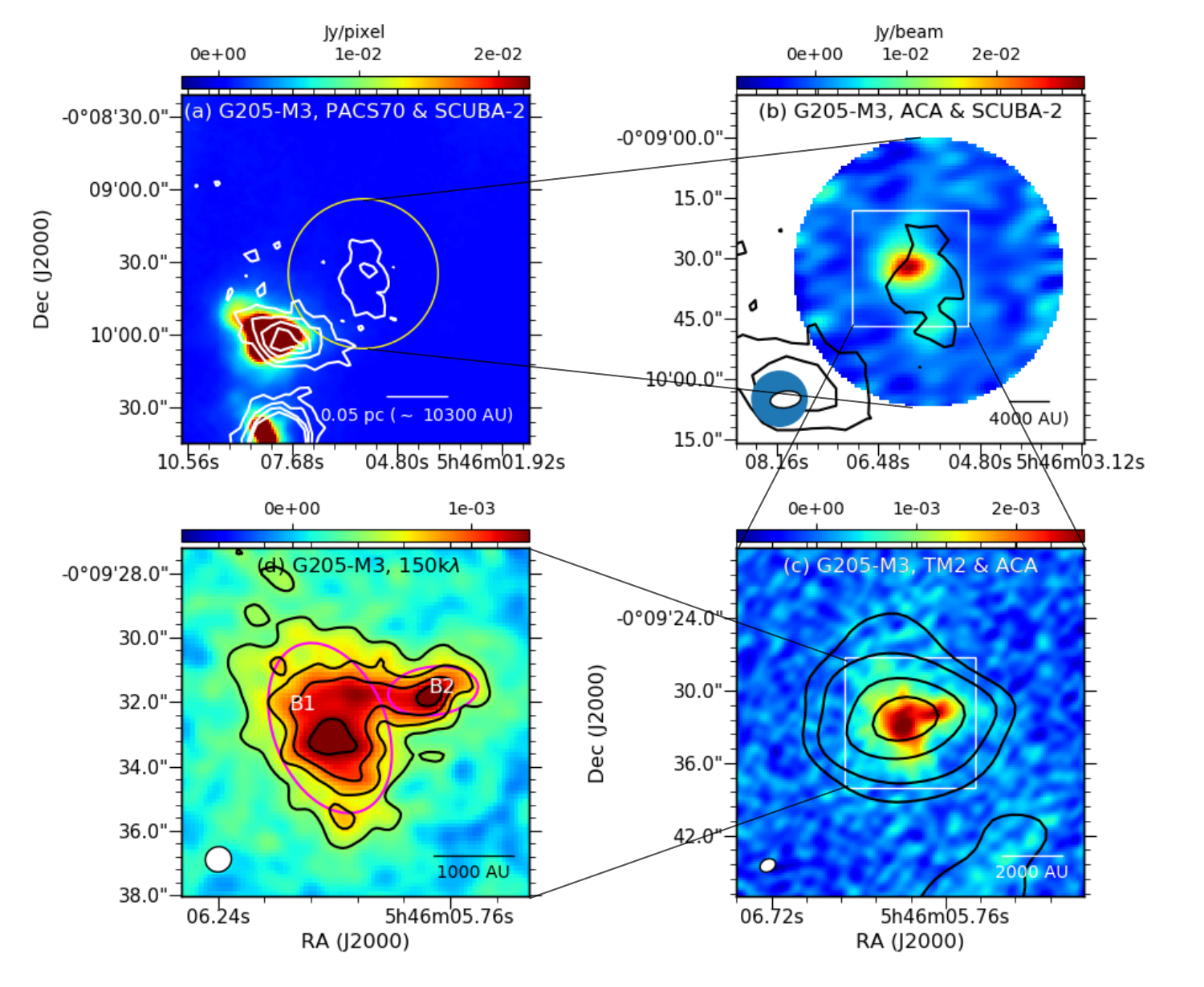}
    \caption{The structure of G205.46-14.56M3 as revealed at different wavelengths and scales. Panel (a) shows 1.3 mm continuum observed by SCUBA-2 in contours:  0.3,0.5,0.7,0.9$\times$peak (where peak = 671.4 mJy/beam corresponds to the  peak flux of closest protostellar core) overplotted on the 70 $\mu$m mid-infrared image taken by Herschel PACS. Panel (b) shows the SCUBA-2 850 $\mu$m continuum in contours (6$\sigma$, $\sigma =38$ mJy/beam) and the ACA 1.3 mm continuum map in color. Both beams are shown in left bottom corner.
    Panel(c) shows the highly dense central region as revealed by the ALMA ACA+TM2 dust continuum map (synthesized beam $\sim 1.2''$). Plotted in contours are the ACA continuum with levels 3,5,10,15$\sigma$.., where $\sigma =1.6$ mJy/beam. Panel (d) zooms toward the central region at a higher resolution (synthesized beam $0.8''$) where the substructures (B1 \& B2) are visible. The contours are at 4,6,10 and 13 $\sigma$ levels  where $\sigma$ = 0.1 mJy/beam. The ellipses correspond to the FWHM sizes inferred from the two-component Gaussian fitting. }
    \label{fig2}
\end{figure*}

We present in Fig.~\ref{fig1} and Fig.~\ref{fig2} (c) the detection of the 1.3~mm (dust) continuum emission in contours toward the five targets (G208.68-19.20N2, G209.29-19.65S1, G209.94-19.52N, G212.10-19.15N1, and G205.46-14.46M3 (G205-M3 hereafter)) with the ALMA ACA+TM2 configurations at a resolution of $\sim$ 1\farcs2. 
As introduced in Section 1, 16 of the 23 candidate starless cores were detected by the ACA observations at a resolution of $\sim$ 6\arcsec  and the 5 targets are among the 16 cores.
Compared with the other 11 detections, the 5 cores discussed in this paper
have further compact dense features within the structures seen by the ACA.
 However, their dust emission is not like the  YSOs,  which have point-like compact emission features as imaged using ALMA-TM1 configuration ($\sim$ 0\farcs3).

\begin{figure*}
    \centering
    \includegraphics[width=18cm]{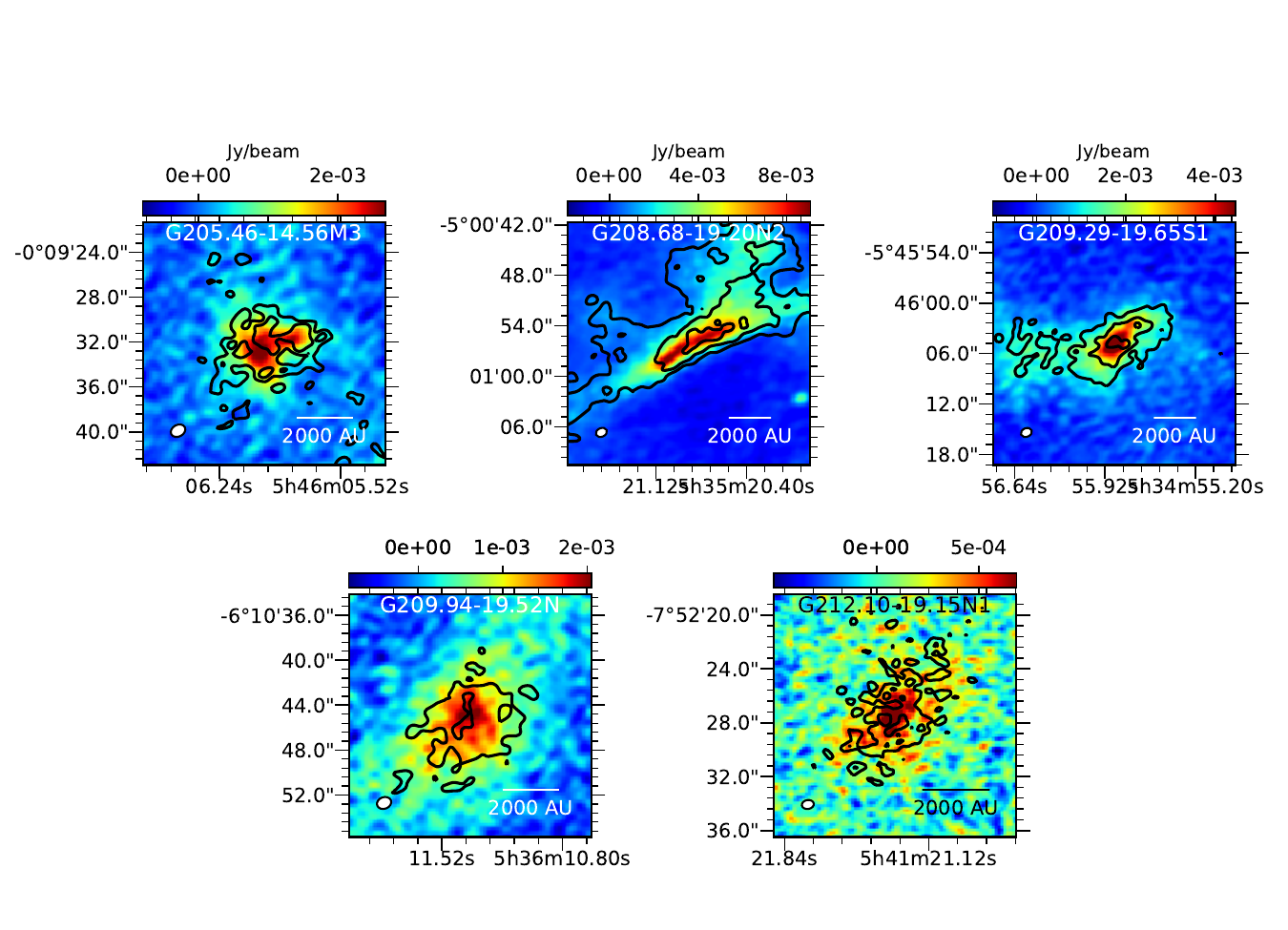}
    \caption{ Integrated \chem{N_2D^+} emission in contours overplotted on the ALMA ACA+TM2 dust continuum map.
    The contours are 3$\sigma$, 5$\sigma$, 7$\sigma$..., where $\sigma$ (rms)=20 and 17 mJy/beam~\kms\ for G205.46-14.56M3 and G212.10-19.15N1 respectively. For G208.68-19.20N2, G209.29-19.65S1, G209.94-19.52N the rms ($\sigma$)=32,20,21 mJy/beam~\kms, respectively and  contours are 5,10,15$\sigma$. } 
    \label{n2dp}
\end{figure*}

None of the five cores displays an outflow signature when observed in CO ($J$=2-1), its isotopologue lines and SiO ($J$=5-4) \citep{Dutta2020}. 
Neither near-infrared nor mid-infrared emission is present within these targets based on our archival search\footnote{\url{(https://irsa.ipac.caltech.edu/irsaviewer/)}}. 
Meanwhile, toward the central region of these five cores, CO appears fully depleted while \chem{N_2D^+} traces well the 1.3 mm (dust) continuum  (see Fig.~\ref{n2dp}). 
This is in accordance with their intense \chem{N_2D^+} emission and high [\chem{N_2D^+}]/[\chem{N_2H^+}] abundance ratio of $\geq 0.1$ seen by the NRO 45m telescope at a larger ($\sim 0.05$ pc to 0.1 pc) scale \citep{Kim2020}.
All these lines of evidence suggest that the five cores are genuinely chemically evolved prestellar cores.  Note that, all of these sources were earlier classified   as `starless' by \citet{Yi2018}, although \citet{Kirk2016} using JCMT survey results classified  G205-M3 as YSO because of the presence of a YSO \citep{Megeath2012,Stutz2013} within the very extended ($\sim 65$\arcsec) envelope surrounding the core. The nearest YSO is  $\sim 27$\arcsec\ away from the position of G205-M3 core (see Fig~\ref{fig2}, panel 1), and therefore not directly associated.

The flux density of the observed structure toward each of the five cores can be estimated from 2-dimensional (2D) Gaussian fitting.
We have neither gas nor dust temperature measurements on the observed angular scales.
Nevertheless, given their prestellar nature, we assume the temperature of the compact component is at most 10~K and can go as low as 6.5~K \citep{crapsi2007,keto2010}.  
 By adopting a (dust) specific absorption coefficient (per mass, with a gas-to-dust mass ratio of 100) $\kappa$ = 0.009 cm$^2$ g$^{-1}$ at the observing wavelength for grains with thick icy mantles after 10$^5$~yr of coagulation \citep{Ossenkopf1994}, we infer the source masses. 
The effective radius of the cores ($R_{\mbox{\scriptsize{core}}}$) is estimated as $\rm{\sqrt{(major \times minor)}/2}$, where `major' and `minor' correspond to the two axes of the ellipse obtained from the 2D Gaussian fitting, and
  the gas column density can be found as $N_{\mbox{\scriptsize H$_2$}} = M/(\pi \mu_{\mbox{\scriptsize H$_2$}} m_{\mbox{\scriptsize H}} R_{\mbox{\scriptsize{core}}}^2 )$.
By assuming a spherical geometry, the gas volume density can also be calculated from  $n_{\mbox{\scriptsize H$_2$}} = 3M/(4\pi \mu_{\mbox{\scriptsize H$_2$}} m_{\mbox{\scriptsize H}} R_{\mbox{\scriptsize{core}}}^3)$, where $ \mu_{\mbox{\scriptsize H$_2$}}$ (=2.8) is the 
molecular weight per hydrogen molecule \citep{Kauffmann2008},  and $m_{\mbox{\scriptsize H}}$ is the proton mass. The mass, size, column density, and volume density of the cores inferred from both the ACA observations and the ACA+TM2 observations 
are presented in Table~\ref{tab:1}. 

\begin{deluxetable*}{ccccccccccc}
\tabletypesize{ \scriptsize}
\tablecaption{ALMA-ACA and ACA+TM2 configuration 1.3 mm continuum results of five presetellar cores   \label{tab:1}}
\tablewidth{2pt}
\tablehead{
\colhead{Source } & \colhead{RA}& \colhead{DEC} &
\colhead{FWHM} & \colhead{ S$_\nu$(1.3 mm)} &
 \colhead{M$_{gas}$} &
\colhead{n$_{H2}$} & \colhead{N$_{H2}$}&  \colhead{Diameter} & \colhead{$L_J$}\tablenotemark{a}  & \colhead{$\alpha$ (FWHM)} \tablenotemark{b}\\
\colhead{} & \colhead{(J2000)}& \colhead{(J2000)} &
\colhead{($\arcsec$)} & \colhead{ (mJy)} &
 \colhead{(M$_{\sun}$)} &
\colhead{(cm$^{-3}$)} & \colhead{(cm$^{-2}$)}&  \colhead{(au)} & \colhead{(au)} & \colhead{ - (\kms)} }
\startdata
\multicolumn{11}{c}{\emph{ACA results}}\\
G205.46-14.56M3* & 05:46:05.99 &	-00:09:32.37 & $8.4\times7.0$ &  79.8 & 0.76-1.69  & $6.4-14.2 \times 10^6$  & $1.9-4.3 \times 10^{23}$ & 3067 & 1561 & 0.68 (0.40)\\
G208.68-19.20N2 &  05:35:20.72 &	-05:00:54.09 & $26.7\times7.7$ & 724.0 & 6.93-15.36 & $8.9-19.7\times10^6$ & $5.1-11.3\times10^{23}$ & 5735 & 1325 & 0.17 (0.49)\\
G209.29-19.65S1 &  05:34:56.04 &	-05:46:05.28	 &$21.5\times7.5$ & 266.0 & 2.55-5.64  & $4.7-10.4\times10^6$  & $2.4-5.4\times 10^{23}$ & 5087 & 1826 & 1.12 (0.97)\\
G209.94-19.52N & 05:36:11.39 &	-06:10:45.93 &$14.2\times 7.2$ & 129.1 & 1.24-2.74 & $4.5- 9.9\times 10^6$ & $1.8-4.0\times10^{23}$  & 4059 & 1868  & 0.83 (0.57)\\
G212.10-19.15N1 & 05:41:21.27 &	-07:52:27.01 &$11.6\times6.0$ & 47.1 & 0.45 - 1.00 & $2.9-6.5\times 10^6$  & $1.0-2.2\times10^{23}$  & 3337 & 2306 & 1.80 (0.55)\\
\hline
\multicolumn{11}{c}{\emph{Combined ACA+TM2 results}} \\
G205.46-14.56M3* &	05:46:05.96 &	-00:09:32.45 & $6.0 \times	4.8$ &	53.2 & 0.51-1.13  & $1.2 -2.8\times 10^7$ &	$2.7 - 5.9 \times 10^{23}$ &	2146 &	1119 & 0.71 (0.40) \\
G208.68-19.20N2	 & 05:35:20.76	 & -05:00:55.21 &	$16.3 \times 3.6$ &	325.0 &	3.11-6.89 &	$2.6-5.7 \times 10^7$	&	$7.9 -17.0 \times 10^{23}$ &	3087 &	781 & 0.21  (0.49)\\
G209.29-19.65S1 &	05:34:55.84 &	-05:46:04.81 &	$7.8 \times 3.9$ &	98.6 &	0.94 - 2.09	& $2.1 -4.7\times 10^7$ &	$4.7-10.0 \times 10^{23}$	 &	2205 &	856 & 1.32 (0.97)\\
G209.94-19.52N	& 05:36:11.38 &	-06:10:45.65	& $10.7 \times	6.5$ &	89.3 & 0.86-1.89 &	$0.6-1.2 \times 10^7$ &	$1.8 - 4.1 \times 10^{23}$ &	3344 &	1680 & 0.99 (0.57)\\
G212.10-19.15N1 &	05:41:21.28 &	-07:52:27.50 &	$7.7 \times 4.2$ &	31.8 & 0.30 - 0.67 & $0.6 - 1.4 \times 10^7$ &	$1.4 - 3.2 \times  10^{23}$ &	2264 &	1568 &1.83 (0.55)\\
\hline 
\multicolumn{11}{c}{\emph{Substructure in G205-M3}}\\
 B1 &05:46:06.008 & -00.09.32.812 &  $5.5\times3.5$  & 39.4  & 0.38-0.84 &   $1.9 - 4.2 \times 10^7$   &  $3.22 - 7.14\times  10^{23}$ &  1755   & 1014  & 0.79 (0.40)  \\
 B2 &05:46:05.795 & -00.09.31.659 & $2.8\times1.5$ &   8.8    &  0.08-0.19  &  $3.7 - 8.2 \times 10^7$ & $3.03 -	6.71\times  10^{23}$   &  820   &  730  &   1.76 (0.40)
\enddata
\tablenotetext{}{ RA and DEC correspond to the peak positions of 2D Gaussian fitting. The range for M$_{gas}$, n$_{H2}$, N$_{H2}$ corresponds to the estimation, assuming \tk= 10~K and 6.5~K respectively.} 
\tablenotetext{*}{The source name in JCMT survey \citep{Yi2018} is G205.46-14.56N1.}
\tablenotetext{a}{Jeans lengths (L$_J$) calculated at 10 K; at 6.5~K the values will be lower}
\tablenotetext{b}{{ Virial parameters calculated} at 10 K with FWHMs obtained by \citet{Kim2020}; at 6.5~K the values will be lower; see section 4.3 for details }
\end{deluxetable*}

\subsection{Substructures in the prestellar core G205-M3}

 G205-M3 is the only core where we detect substructures at a scale of 1000 au inside the compact dense structure of the core  (also see section 4.1.1).
We show in Fig.~2 the (dust) continuum images of the G205-M3 core.
Different $uv$--tapering of the visibility data has been employed to highlight the intricate features of the core seen at different angular scales.
The G205-M3 core size in SCUBA-2 observation (beam size $\sim 14''$) is on the order of 0.05 pc or 10000 au \citep{Yi2018}.
The core is detected by ACA (beam $\sim 6\arcsec$, see Fig. 2(b)) with an overall size on the order of 4000 au and moreover, the faint substructure can be readily discerned.
At a higher (1\farcs2) resolution (see Fig. 2(c)) an asymmetric structure can be seen with an average core size of $\sim 2000$ au. 
Further zooming into the source with an angular resolution of 0\farcs8 (
corresponding to a uv-taper of 150~k$\lambda$; Fig.~2-d), the inner  2000~au region of G205-M3 is then resolved into two noticeable substructures. 
These substructures are named B1 and B2 ( see Fig~\ref{fig2} -panel d). 
These substructures were also  identified using a dendrogram analysis, see Fig.~\ref{dendrogram}.
They were fitted simultaneously with 2D Gaussian fitting. 
The two components are, respectively, 5\farcs5 $\times$ 3\farcs5\ and 2\farcs8 $\times$ 1\farcs5 in size and separated roughly by 1200 au (3\arcsec).
The masses of B1 and B2 are 0.38-0.84 M$_\odot$ and 0.08-0.19 M$_\odot$, respectively, if a (dust) temperature range of 6.5-10~K is assumed, with larger masses associated with the lower temperatures. 
Their corresponding volume densities, considering a spherical geometry, are listed in Table~\ref{tab:1}.

\section{Discussion}

 Summarizing the results, a series of observations of increasingly better resolution and sensitivity has revealed increasingly compact, denser structures in a small subset of starless cores. In this section, we discuss the physical properties, the stability, and the dynamical state of these structures.

\subsection{Detection of  compact structures}
In the ACA/ALMA maps, the  flux densities of G205-M3 enclosed in apertures with an equivalent radius
$r$ of 0\farcs4, 1\farcs1, 2\farcs2,  5\farcs7, and 14\farcs9 centered at the peak position of B1 are 1.4~mJy, 7.5~mJy, 24.8~mJy,  52.2~mJy,and 99.5~mJy respectively. Assuming the gas temperature and dust property are uniform within these scales, we find the corresponding enclosed masses will scale as $r^{1.2}$, which hints at a density profile of $r^{-1.8}$ between 300~au and 10000~au scale. 
Using similar method, for other cores (except G208-N2) the masses scale as $r^{1.2} - r^{1.3}$, and therefore their density profiles vary as 
$r^{-1.7} - r^{-1.8}$.
The actual density profile could be even steeper if the dust temperature decreases toward the center of the core.
These structures are thus not consistent with Bonnor-Ebert spheres unless the inner region of constant density is very small \citep{Dunham2016}.
Such systems are generally very close to the initiation of collapse or they have begun to collapse.
Based on their statistical analysis on the lifetime of starless cores, \citet{Dunham2016}  and also \citet{Kirk2017} argued that only evolved cores with short lifetime ($\sim 10^4$ year) can be detected by interferometers. Therefore, the five cores detected at high resolution ($\sim$ 1\farcs2) likely represent an evolved starless stage, just before the onset of star formation. 
 
 Such dense regions can be compared with the well known prestellar core `kernel' in L1544 \citep{Caselli2019}.
The average and peak densities of the five prestellar cores are at least 10 times higher than that of the L1544 core, although the sizes of the regions are comparable.

 \subsubsection{Search for substructures at  1000~au scale}
 Do the five cores have substructures inside the compact dense regions?
 We use different methods, e.g., uv-tapering of  the image, slicing along possible substructure features to 
 identify the intensity variation, and the dendrogram technique. This helps us identify the real and prominent substructures which are not related to 
 imaging artifacts \citep[e.g.,][]{Caselli2019} and weak substructure features. Prominent signatures
 of substructures are  found only towards G205-M3 (section 3.2). Among the other four cores, the G208-N2 
 core appears elongated and similar to a filamentary structure (More details of this core will be presented 
 by Hirano et al. in preparation). 
   However, the intensity fluctuations along the apparent substructures are at most comparable to the noise level, therefore we see no evident signs of substructure towards G208-N2. Similarly, we do not find 
 substructures toward the other remaining cores, and their detailed physical evolutionary status and density structure will be described in a future work (Sahu et al. in preparation).

 \subsection{Thermal stability}
 
Assuming first an isothermal molecular core, being supported only by its thermal pressure against gravity without considering non-thermal gas motions, one can estimate the Jeans length scale beyond which gravitational collapse would prevail.
The Jeans length is defined by: $ L_J=\sqrt{\frac{\pi {C_s}^2}{G \rho_0}}$,
where $G$ is the gravitational constant, $\rho_0$ is mass density and \chem{C_s}  is the isothermal sound speed, \chem{C_s = (k_B T/\mu_p m_H)^{0.5}}, with $\mu_p = 2.37$ \citep{Kauffmann2008}. 
\citet{Yi2018} found that at the scale of the SCUBA-2 observation, these five cores are unstable based on the Jeans analysis. 
We find similar results for the dense structures seen at the ACA (6$''$) and TM2($\sim 1.2''$ scales (see Table~\ref{tab:1}). 
The dense structures in the five cores have sizes significantly greater than their correspondingly Jeans lengths, implying that they are Jeans unstable.

The adopted gas temperature is a major uncertainty in the above analysis.
While we have adopted a gas temperature of 10~K for all sources, \citet{Kim2020} found from \chem{N_2H^+} and \chem{N_2D^+} line observations that under LTE conditions the gas temperatures of the five cores are in the range of 10.8-17.3~K at  larger scales with an average density ($\sim 10^5$\cmq).
Nevertheless, the cores remain unstable even if we adopt a gas temperature of 20~K.
In fact, 
the temperatures of the central compact dense regions are likely to be lower than 10~K given their high density.
A good example is L1544, for which \citet{Caselli2019} assumed a core temperature to be 6.5 K for its high density ($\sim 10^6$\cmq).
The average density of the central dense regions in our core sample is at least ten times higher than L1544 and a gas temperature of 6.5 K is therefore viable. 
At this temperature all the core structures detected by ACA are Jeans unstable, further highlighting their prestellar nature.

\subsection{Dynamical state}

 We particularly focus on the dynamical state towards the G205-M3 core, where substructures are detected. 
 To assess the dynamical state of the regions, we use molecular line data to determine thermal and non-thermal contributions, using \chem{{\sigma_v}^2={\sigma_{th}}^2 +{\sigma_{nt}^2}}, where \chem{\sigma_{th} = (k_B T/m)^{0.5}}, with $m$ the mass of the observed molecule, and \chem{\sigma_{nt}} are  the thermal and non-thermal components, respectively. The non-thermal contributions could include infall, rotation, and streaming motions, but they are usually attributed to turbulence. Ideally, we would use the line data on the scales of the substructures, but our velocity resolution is insufficient. We instead consider information from NH$_3$ with an FWHM linewidth (\dv) of 0.88 \kms \citep{Cesaroni1994}
on a scale of 40 arcseconds ($2\ee4$ au)
and from \ntdp\ with $\dv = 0.4$ \kms\ on a scale of 8000 au. Using $\dv = (8 ln 2 )^{0.5}\rm{\sigma_v}$, we obtain \chem{\sigma_v = 0.37} \kms\ for NH$_3$ and 0.17 \kms\ for \ntdp. If we assume $\tk = 10$ K, \chem{\sigma_{th} = 0.068} \kms\ for NH$_3$ and 0.053 \kms\ for \ntdp, resulting in \chem{\sigma_{nt} = 0.37} \kms\ for NH$_3$ or \chem{\sigma_{nt} = 0.16} for \ntdp, with the smaller value more likely relevant to the individual cores and indicating nearly equal contributions from turbulence and thermal broadening. The isothermal sound speed, \chem{C_s} 
is 0.19 \kms\ at $\tk = 10$ K, indicating that the turbulence is supersonic,
with Mach number of 2 based on NH$_3$ or transonic, with Mach number of 0.8 based on \ntdp. Taken at face value, these results suggest that the turbulence has decayed on small scales. The effective sound speed,  \chem{c_{s,eff} = (C_{ s}^2 + \sigma_{nt}^2)^{0.5}}, is 0.247 \kms, using the data from \ntdp.

The fate of the substructures can be assessed from their virial parameters. The parameter $\alpha_{\rm vir} = 2a E_{\rm k}/|E_{\rm g}|$. where $E_{\rm k}$ is the kinetic energy and $E_{\rm g}$ is the gravitational potential energy, and $a = 2\pm1$ for a wide range of geometries and density structures \citep{Kauffmann2013}.
The virial parameter can be written as $\alpha_{\rm vir} = 5.6\ee{-3} a c_{\rm s, eff}^2 R(\rm au) M(\msun)^{-1}$. Taking the sizes and masses and the effective sound speed of 0.247 \kms, we obtain $\alpha_{\rm vir} = 0.36a$ to $0.79a$ for B1 and  $\alpha_{\rm vir} = 0.74a$ to $1.76a$ for B2. This calculation suggests that the substructures are close to being gravitationally bound, even if all the non-thermal component is entirely due to turbulence. Other motions, such as infall, may be present. It is at least plausible that the sub-structures are collapsing to form separate objects. Using a similar calculations, we found that the four other dense cores are also close to being gravitationally bound (see Table~\ref{tab:1}). This again supports the evolved prestellar nature of the cores. 

\subsection{Substructures toward G205-M3 and stellar multiplicity}
 We have detected,  for the first time, substructures within the central compact dense region ($\sim$ 2000 au) of the prestellar core G205-M3. What will be the fate of these kinds of substructures? 

If the substructures are in free-fall, they are likely to form separate objects that could be members of a wide binary. They are unlikely
to coalesce before they collapse. The minimum time to coalesce is the crossing time. Using the projected separation of 1148 au ( also see the appendix section) and the effective sound speed of 0.247 \kms, we calculate $t_{cross} = 5.0$ yr $s/c_{\rm s, eff} = 2.3\ee4 $ yr, where $s$ is the projected separation in au and $c_{\rm s, eff}$ is in \kms. This estimate assumes that there is no separation along the line of sight and that they are headed on a collision course. In contrast, the free- fall time is
 $\sqrt{3\pi/32 \rho_0 G }$ (where, $\rho_0=n_{H_2}\mu_{H_2}m_{H}$ with $\mu_{H_2}$=2.8), for a minimum H$_2$ number density (n$_{H_2}$) of 1.9\ee7 \cmq, $t_{\rm ff} < 7.06\ee3$ yr, which is much shorter than $t_{cross}$. On the other hand, a binary system will be bound if the internal energy: $E_{int}=1/2\mu v^2 - G m_1 m_2/r$ $< 0$, where $\mu$ is the reduced mass of the two bodies. Considering the mass of B1 and B2 substructures/sub-cores, the system will be bound for a velocity difference ($v$) of 0.8-1.2 \kms. From the low resolution spectral data ( \chem{N_2D^+}), we find that the  velocity difference is at  most 1.465 \kms. Therefore, it is very plausible that the   sub-cores may eventually form  a wide binary system.  Notably, \citet{Karnath2020} reported substructures associated with protostellar candidates. Those substructures appear to trace an early stage of protostellar evolution, during which substructures are associated with collapsing fragments and individual components may be optically thick hydrostatic cores. The detected structures toward the prestellar core (G205-M3) are therefore not some transient features; they likely persist into the protostellar phase as well.

\subsection{Physical explanations for fragmentations}

 In the `gravoturbulent fragmentation’ \citep{Palau2018} scenario fragmentation takes  place in a self-gravitating turbulent medium.  In this case, the density is determined by enhancements created by turbulence \citep[e.g.,][]{Padoan2002,Fisher2004,Goodwin2004,Offner2010}. Based on this theory, \citet{Offner2012} predicted that prestellar core fragmentation can be observable at 1000 au scale. It is plausible that fragmentation towards the G205-M3 core represents such a case.
 However, the non-detections of sub-cores in other prestellar cores possibly imply a younger stage than the G205-M3 core, and they may fragment in a later period of their evolution.
 
 \section{Summary}

We present 1.3 mm dust continuum observations using different configurations of ALMA, resulting in different synthesized beams  to study five highly dense ($> 10^7$\cmq) prestellar cores in the Orion molecular cloud. We found that in addition to detection using the ALMA-ACA-configuration ($6''$), these cores are also detected using the ALMA-TM2-configuration ($1.2''$) which imply that the cores have a centrally dense region of size $\sim 2000$ au. 
No  NIR/MIR emission  has been detected towards these cores,  signifying that the cores are starless/prestellar in nature. 
The cores are found to be gravitationally unstable, and at the onset of star formation
We found two substructures of sizes ranging from 800-1700 au and masses $ >0.08-0.84$ M$_\odot$ towards the core G205-M3.  Considering that the free fall time is shorter than the coalescence time of the substructures, and they are likely bound within a  separation of $\sim$ 1200 au, we speculate 
 that this core will produce a wide binary or multiple star system.

\acknowledgments
This paper makes use of the following ALMA data: ADS/JAO.ALMA\#2018.1.00302.S. ALMA is a partnership of ESO (representing its member states), NSF (USA) and NINS (Japan), together with NRC (Canada), MOST and ASIAA (Taiwan), and KASI (Republic of Korea), in cooperation with the Republic of Chile. The Joint ALMA Observatory is operated by ESO, AUI/NRAO and NAOJ.
We thank the reviewer for his helpful comments and suggestions. DS and SYL acknowledge support from the Ministry of Science and Technology (MoST) with grants 108-2112-M-001-048- and 108-2112-M-001-052-. Tie Liu acknowledges the supports from the international partnership program of Chinese academy of sciences through grant No.114231KYSB20200009, National Natural Science Foundation of China (NSFC) through grant NSFC No.12073061, and Shanghai Pujiang Program 20PJ1415500. 
N.H. acknowledges MoST 108-2112-M-001-017, MoST 109-2112-M-001-023 grant.
G.G. acknowledges support from ANID project AFB 170002.
L.B. acknowledges support from CONICYT project Basal AFB-170002.
S.L. Qin is supported by National Natural Science Foundation of China under grant No. U1631237. DJ is supported by NRC Canada and by an NSERC Discovery Grant. 
VMP acknowledges support by the Spanish MINECO under project AYA2017-88754-P, and financial support from the State Agency for Research of the Spanish Ministry of Science and Innovation through the ``Unit of Excellence María de Maeztu 2020-2023'' award to the Institute of Cosmos Sciences (CEX2019-000918-M). C.W.L. is supported by the Basic Science Research Program through the National Research Foundation of Korea
(NRF) funded by the Ministry of Education, Science and Technology (NRF-2019R1A2C1010851). A.S. acknowledges financial support from the NSF through grant AST-1715876.
The research was carried out in part at the Jet Propulsion Laboratory which is operated for NASA by the California Institute of Technology.

\bibliography{orion_p1.bib}{}
\bibliographystyle{aasjournal}

\appendix
\restartappendixnumbering
\section{Dendrogram analysis}
\begin{figure}
    \centering
    \includegraphics[width=0.5\textwidth]{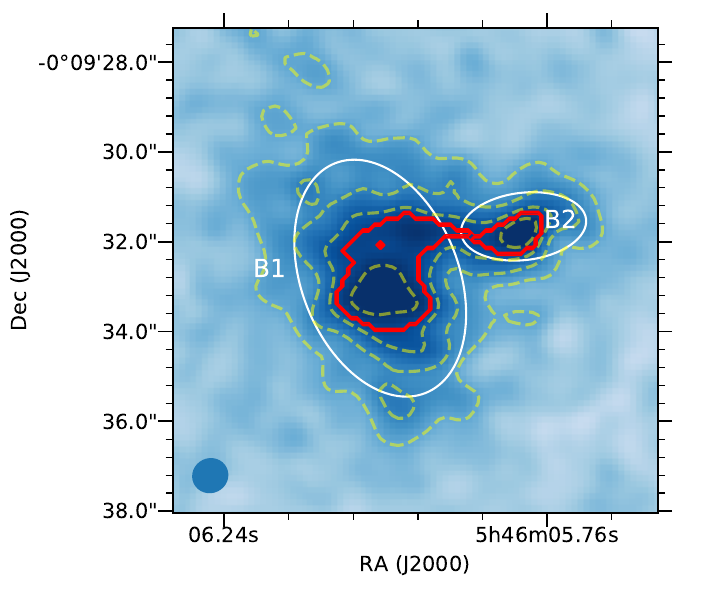}
    \caption{ The red-contours show the condensations/blobs as identified using dendrogram analysis.  White ellipses corresponds to FWHM size of the two component Gaussian fitting of the condensations B1 and B2. Dotted contours corresponds to 1.3mm dust continuum at 0.8$''$ resolution with the levels 4,7,10,13$\sigma$   where the rms noise is given by $\sigma=0.1$  mJy/beam.}
    \label{dendrogram}
\end{figure}

In addition to visual identification of substructures towards the prestellar core G205-M3 (see Fig~\ref{fig2}, panel d),
we also show  the identification of the substructures with the dendrogram analysis, which is often employed in identifying and labeling clumpy structures \citep{Rosolowsky2008,Pineda2015}.
We ran the dendrogram algorithm with min\_value =4$\sigma$ (minimum intensity considered in the analysis) and min\_delta= 2$\sigma$(minimum spacing between isocontours). 
Two substructures are clearly identified and presented in Fig.~\ref{dendrogram} (red contours).  The separation between the blobs, B1 \& B2 as obtained from the dendrogram analysis, based on the peak position of the blobs is 1148 AU. If we consider the position of the peaks of B1 and B2 as obtained from Gaussian fittings (presented in Table~\ref{tab:1}) then the separation is given by 1360 $\pm$ 140 AU, considering the fitting uncertainties. So, in general we consider the separation $\sim$ 1200 AU, as mentioned in the text.

The image presented in Fig.~\ref{dendrogram} corresponds to the combined data of ALMA (ACA, TM2 \& TM1 configurations) with a uv-taper of 150~k$\lambda$ ( $\sim$0\farcs8). The dust emission of the five dense  prestellar cores which were detected in ALMA-TM2 ($\sim$1.2\arcsec), resolved out at the highest observing resolution (ALMA-TM1; $\sim$ 0\farcs3). So, to identify the substructures from the dust continuum images, we applied a range of uv-tapering on the combined ALMA data from 100~k$\lambda$ to 200~k$\lambda$ (corresponds to resolution from 0\farcs6 to 1.0\arcsec). The substructures of G205.46-14.56M3 appear most prominent
in ALMA combined data at $\sim$0\farcs8 - 1.0\arcsec resolution (with uv-taper of 100 - 150~k$\lambda$), but at higher resolutions (beyond 200~k$\lambda$) their emission was resolved out. 




\end{document}